\documentclass[aps,pre,preprint,groupedaddress,showpacs]{revtex4}

\usepackage{epsfig,color,times}
\usepackage{amsmath, mathrsfs}
\begin{document}


\title{Gravity driven instability in elastic solids}

\author{Serge Mora}
\email[]{smora@univ-montp2.fr}
\affiliation{Laboratoire de M\'ecanique et de G\'enie Civil de Montpellier. UMR 5508, Universit\'e Montpellier 2 and CNRS. Place Eug\`ene Bataillon. F-34095 Montpellier Cedex, France.}

\author{Ty Phou}
\affiliation{Laboratoire Charles Coulomb. UMR 5521, Universit\'e Montpellier 2 and CNRS. Place Eug\`ene Bataillon. F-34095 Montpellier Cedex, France.}

\author{Jean-Marc Fromental}
\affiliation{Laboratoire Charles Coulomb. UMR 5521, Universit\'e Montpellier 2 and CNRS. Place Eug\`ene Bataillon. F-34095 Montpellier Cedex, France.}

\author{Yves Pomeau}
\affiliation{University of Arizona, Department of Mathematics, Tucson, USA.}

\date{\today}
\begin{abstract}
We demonstrate the instability of the free surface of a soft elastic solid facing downwards. Experiments are carried out using a  gel of constant density $\rho$, shear modulus $\mu$, put in a rigid cylindrical dish of depth $h$. When turned upside down, the free surface of the gel undergoes a normal outgoing acceleration $g$.  It remains perfectly flat for $\rho g h/\mu<\alpha^* $ with $\alpha^*\simeq 6$,  whereas a steady pattern spontaneously appears in the opposite case. This phenomenon results from the interplay between the gravitational energy and the elastic energy of deformation, which reduces the Rayleigh waves celerity and vanishes it at the threshold. 
\end{abstract}

\maketitle
\begin{figure}[!h]
\begin{center}
\includegraphics[width=0.5\textwidth]{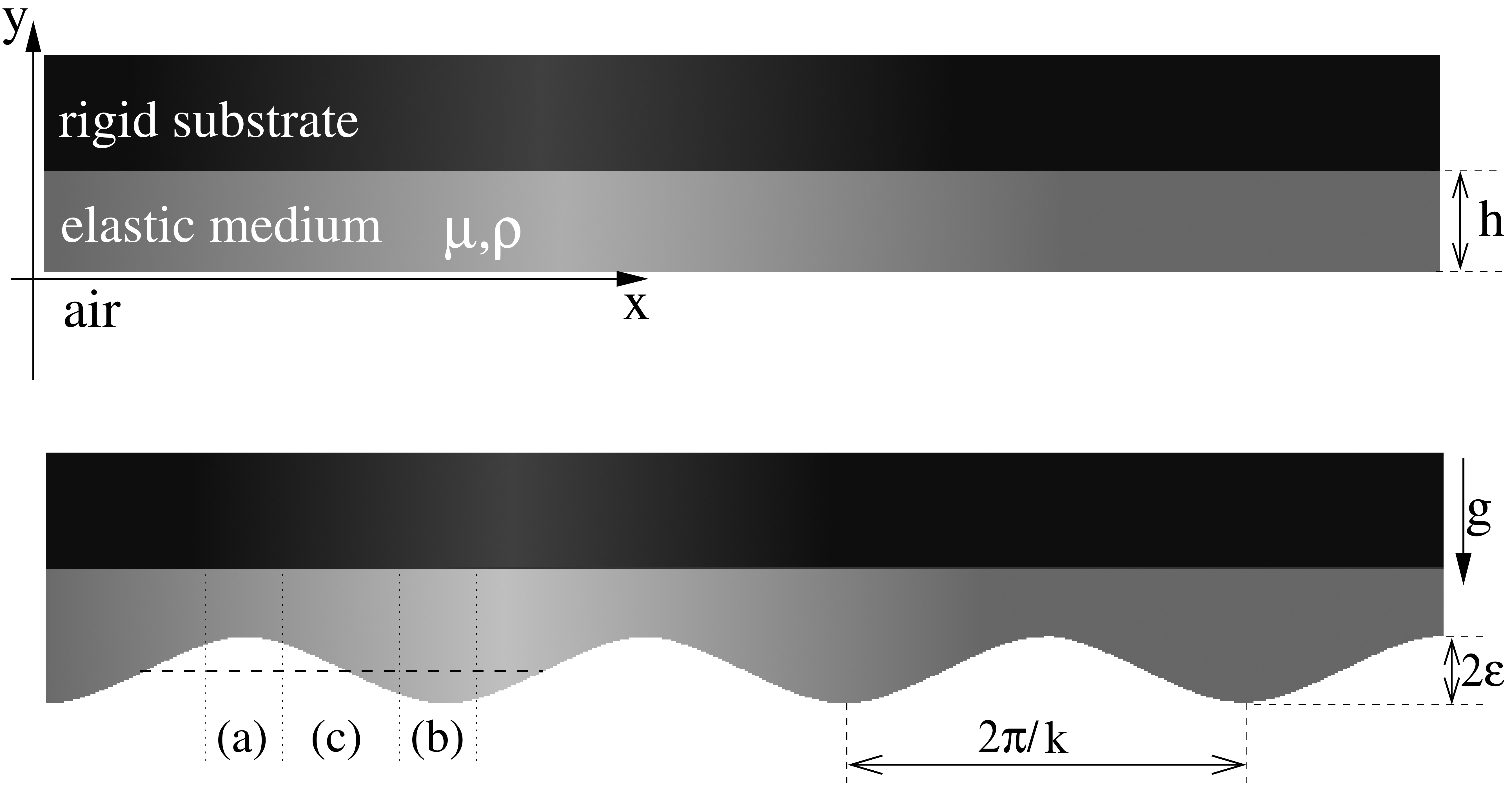}
\end{center}
\caption{Scheme of a sinusoidal disturbance of a downwardly facing and initially flat surface of a heavy elastic slab. The other surface is fixed on the rigid substrate. The energy change due to this disturbance can be positive or negative depending on the value of the dimensionless ratio $\rho g h/\mu$ with $\rho$ the mass density, $g$ the (gravitational) acceleration, $h$ the thickness and $\mu$ the elastic modulus.} \label{fig : qualitative}
\end{figure}

 Many materials such as biological tissues can withstand huge elastic deformations of more than several hundred percent. The amplitude of the stress is then of the order of the elastic modulus, a situation commonly encountered  with soft materials.  Specific and fascinating  patterns, reminiscent of those that can be seen in hydrodynamics, can then occur spontaneously \cite{Shull2000, Adda2006, Mora_prl2010,Mora_softmatter2011,Dervaux2012, Style2013, Saintyves2013, Biggins2013, Tallinen2013, Mora_prl2013}. 
Since both soft elastic solids and liquids are capable of undergoing large deformations, and are often subjected to forces with a common origin, eg capillary forces \cite{Style2013prl,Style2013}, it is likely that some mechanical instabilities can be shared, to a certain extent, by these two kinds of continuous media \cite{ Mora_prl2010}. 
Out of the many instabilities experienced by liquids, the Rayleigh-Taylor instability (RTI)\cite{Rayleigh1883,Taylor1950,Sharp1984} is outstanding because it is easy to understand, not too difficult to rationalize and also important in many technological and physical situations.
The dispersion relation for regular gravity waves on a deep ocean reads $\omega = \sqrt{ g k}$, where $g$ is the downward gravity acceleration, $\omega/(2 \pi)$ is the wave frequency and $k$ its horizontal wave number. As often noticed, if one turns the gravity upward, that is if one changes the sign of $g$, $\omega$ becomes purely imaginary $\pm i \sqrt{-g k}$, showing the existence of fluctuations growing exponentially with time. These fluctuations do not saturate and yield ultimately fingers of liquids in free fall. If one considers, as we do below, a soft solid in air with its surface turned downward, there are a priori good reasons to believe that some sort of RTI will set in.
To figure it, consider a horizontal elastic slab of thickness $h$ subjected to the gravity of Earth $g$, the upper surface being fixed to a rigid body, the lower one being free  (Fig.\ref{fig : qualitative}). A sinusoidal perturbation $\zeta=\varepsilon \sin(kx)$ of the surface height (with $x$ an in-plane coordinate) causes a reduction in the gravitational energy per unit area equal to $\frac{1}{2\lambda}\int_{0}^{\lambda} \rho g \zeta^2\mathrm{d}x=\frac{1}{4}\rho g \varepsilon^2$, where $\rho$ is the mass density of the elastic medium and $\lambda=2\pi/k$ is the wave-length. The corresponding elastic energy cost per unit volume scales as the shear modulus $\mu$ times the mean squared strain. In the long wave limit ($kh\ll 1$) the strain scales as $\epsilon/h$: the sample is vertically squeezed from length $h$ to $h-\varepsilon$ above a trough of the wave (region (a) of  Fig.\ref{fig : qualitative}), it is vertically stretched above a peak from $h$ to $h+\varepsilon$  (region (b)), and the deformation varies progressively in between (region (c)). Finally, the mean elastic energy per unit volume scales as $h\cdot \mu (\varepsilon/h)^2$.  Comparing the two contributions of the total energy, it appears that the Rayleigh-Taylor buoyancy overcomes elasticity beyond an instability threshold  $\rho g h/\mu=\alpha^*$ where $\alpha^*$, the dimensionless proportionality constant for the elastic energy (with a factor of four), is to be found (See Supplement-B for the complete calculation). \\
In the same situation a thin layer of liquid is always unstable  which is equivalent to set to zero the shear modulus in the previous estimate. Therefore, contrary to liquids  RTI in a solid has a well defined threshold for layers of finite thickness. Beyond it, the deformation increases up to a finite value for which the elastic cost balances the buoyancy gain: a steady state of equilibrium is then reached. \\

Although RTI in solids is expected to play  a role in many fields such as biology, geology \cite{Houseman1997,Burov2008} or astrophysics \cite{Blaes1990}, both a direct observation and a clear characterization are missing. In \cite{Barnes1974,Barnes1980} and \cite{Lebedev1996}, a flat metal plate whose thickness is initially periodically modulated with a low amplitude is accelerated by expanding detonation products. The growth of the initial perturbation is observed through the use of x-ray shadow-graphs. It was found to be governed by the yield strength of the elasto-plastic material, the initial amplitude and the plate thickness. More recently, yogurts  with a sinusoidal perturbation at the surface were put in a mold and accelerated using a linear electric motor.  The stability regions of this elastic-plastic material have been investigated in terms of acceleration, amplitude and wavelength of the initial perturbation \cite{Dimonte1998}. In both cases (flat metal plate and yogurt), the observations consist in evolving states which are clearly associated to {\em plastic} deformations of {\em pre-existing} periodic ripples at the free surface. Schematically, when the acceleration exerts a strong enough stress on the ripples, the yield stress of the material is overcome at the ripples extremities which begin to flow. In the experiments of \cite{Dimonte1998}, the case of an initially flat surface has been briefly investigated. The authors reported the existence of a non stationary surface instability and related its nucleation to the elastic (reversible) deformations of the material. Their conclusion seems erroneous insofar it is based on a comparison with a theoretical expression valid for samples whose height is much higher than the wavelength, which is not the case in their experiments. The results obtained in the present paper demonstrates unambiguously that if the phenomenon observed by these authors were a consequence of  RTI for an elastic solid, their observations would have been different. It is therefore likely that the observed phenomenon is a consequence of the plastic properties of the investigated material. 

Following the pioneering analytic work of Drucker \cite{Drucker1980},  RTI in {\em plastic} solids has been modeled in the visco-elasto-plastic approximation in order to simulate the growth in amplitude of initial sinusoidal perturbations \cite{Miles1966,Swegle1989,Abakumov1990}. RTI for purely elastic plates with an initially flat surface has been analytically studied by Plohr and Sharp \cite{Plohr1998}, whose results have been generalized few years after \cite{Terrones2005,Piriz2009}. They predict for each value of the acceleration the existence of a critical perturbation wave-length beyond which the flat surface is unstable. As a consequence, an elastic plate is always unstable, provided its dimension are large enough compared to the unstable wavelengths.  This is in contrast with the instability studied in this letter. On the other hand, Bakhrakh and Kovalev \cite{Bakhrah1978} have analytically studied the case of an accelerated elastic half space and found that it is unstable with respect to any perturbations with a wavelength larger than $4\pi\mu/(\rho g)$. \\      
 
We report below the experimental observation of an instability occurring on the surface of a heavy ideal elastic solid pointing downwards. This instability occurs above a threshold and results in steady patterns. At threshold, elasticity exactly counterbalance buoyancy for an infinitesimal perturbation of the free surface. This phenomenon is closely related to Rayleigh waves \cite{Rayleigh1885} since the phase velocity of elastic surface waves decreases as the inwardly gravity increases. This provides a physical interpretation of the instability we have demonstrated, insofar it occurs when the gravity is strong enough to make fully vanish the phase velocity. This viewpoint leads us straightforwardly to calculate the growth rate of the instability.  \\

In our experiments, we use aqueous polyacrylamide gels consisting in a loose permanent polymer network immersed in water. The density of this incompressible elastic material is almost equal to that of water. It behaves as an elastic solid for strains up to several hundreds of percent (Supplement-A). The shear modulus can be tuned over a wide range by varying the concentrations in monomers and cross-linkers, or the temperature. In our experiments, it lies between 30 and 150 Pa. It is measured through indentation tests (Supplement-A).\\

\begin{figure}[!h]
\includegraphics[width=0.5\textwidth]{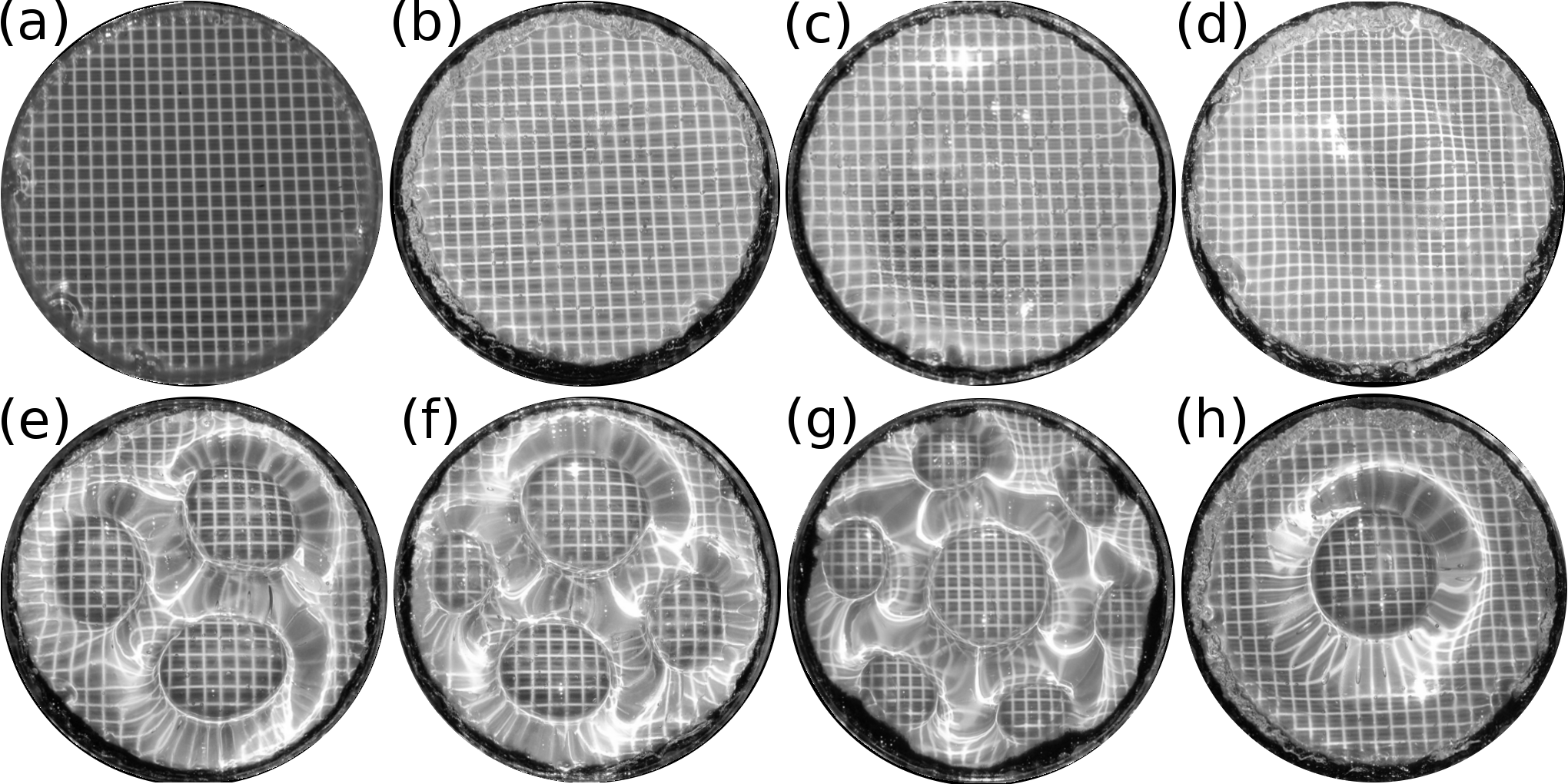}
\caption{Views of the downwardly facing free surface of gels with different shear moduli. (a) $\mu=78 \pm$ 0.5 Pa; (b) 44  $\pm$ 0.5 Pa; (c) 43.3 $\pm$ 0.5 Pa; (d) 42.8  $\pm$ 0.5 Pa; (e) 41.0 $\pm$ 0.5  Pa; (f,g,h)  40.0 $\pm$ 0.5 Pa.  The cylindrical dish is 18 cm diameter and 2.75 cm deep. (a-g) Steady patterns obtained just after reversal.  (h) The sample temperature was 50 degrees when reversed. The snapshot is taken after the sample has cooled to room temperature.}
\label{fig : module}
\end{figure}
The reagents generating the gel are dissolved in ultra-pure water and poured into the brim in a cylindrical dish whose walls are covered with a thin layer of Velcro loops to prevent any further detachment. After the gel is made and its shear modulus measured, the dish is flipped upside down. Various methods have been tested: (i) reversal when the system is immersed in water (density close to that of the gel), the container is then gently removed out of the water keeping horizontal the free surface; (ii) reversal carried out in air but with a rigid plate keeping flat the surface during inversion. The plate is then gently removed; (iii) direct and fast flipping of the system in air without any special care. The three methods lead to identical results. The surface of the thinnest or the hardest samples remains perfectly flat (Fig.\ref{fig : module}-a).  In a narrow range of shear moduli and heights non propagating undulations grow spontaneously at the free surface of the gel and remain permanently (Fig.\ref{fig : module}-b,c,d). For lower shear moduli or for greater thicknesses, several cuvettes appear next to each other at the surface, and remain permanently. For a constant thickness, their number (from one to seven in our experiments) and their size depend on the shear modulus (Fig.\ref{fig : module}-e,f,g). In any cases, flipping again the container (so that the free surface is horizontal and upward) leads to the perfectly flat surface we started from. In addition, successive reversals lead to the same observations, except for a particular sample for which the number of cuvettes is either four, either seven (Fig.\ref{fig : module}-f,g). We infer that the shear modulus of this sample corresponds to a threshold for the number of cuvettes so that the final configuration of the system is driven by uncontrolled external disturbances. This point is discussed at the end.

\begin{figure}[!h]
\includegraphics[width=0.5\textwidth]{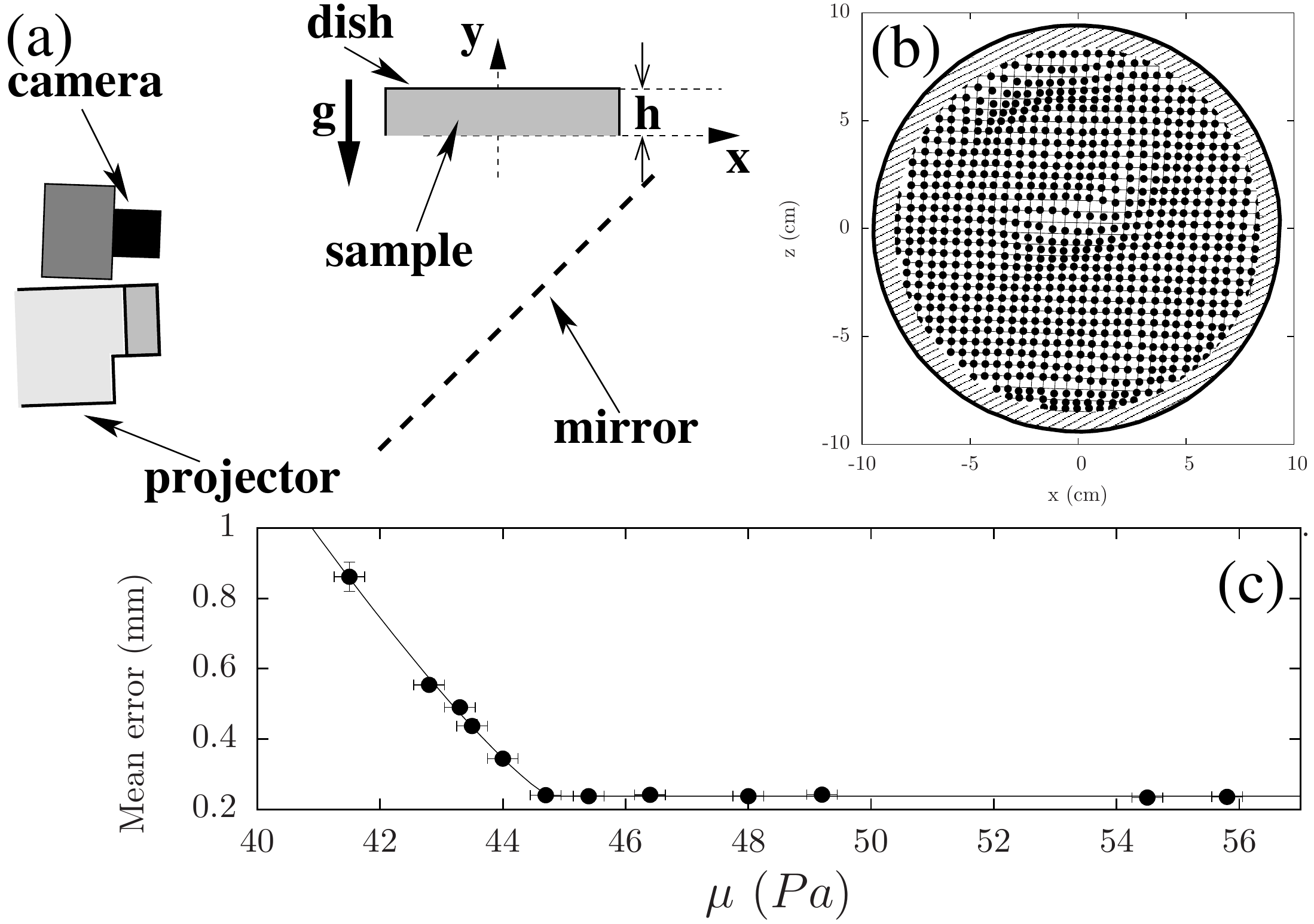}
\caption{Quantitative analysis of the surface distortion. {\bf (a)} Experimental setup. The distance mirror-projector-sample is 1.5 m while the dish diameter is 18 cm. {\bf (b)} Full circles: intersections of the distorted lines of the grid observed on a sample ($\mu= 41.5 \pm 0.5$ Pa). Solid lines defined the grid that best fits the observed one. {\bf (c)} Square root of the mean squared error with the grid that best fits the observed one as a function of $\mu$ with $h=2.75$ cm  (full circles). The solid line is the best fit with the power law $a+b(\mu^*-\mu)^c$. We find $\mu^*=44.6\pm 1.8$ Pa. }
\label{fig : fit grid}
\end{figure}

To obtain quantitative information about the surface deformation, a regular light grid is projected about the free surface (Fig.\ref{fig : fit grid}-a). The gel being transparent and the bottom of the container being white, the observed image of the grid results from one refraction followed by one reflection and another refraction.  If the free surface is flat, this image corresponds to the grid without geometric distortion (Fig.\ref{fig : module}-a). It is warped if the free surface is deformed, this distortion is the bigger the surface is more deformed (Fig.\ref{fig : module}-b-g). To measure the distortion, a rectangular lattice is fitted with the recorded images using the least squares method. The fitting parameters are a translation, the parameters and the orientation of the lattice, and a possible quadratic distortion taking into account the  (small) radial decentering optical distortion (Fig.\ref{fig : fit grid}-b). Fig.\ref{fig : fit grid}-c shows this deviation plotted as a function of $\mu$ for samples having the same thickness $h=2.75$ cm. Fitting functions $a+b(\mu-\mu^*)^c$ with these data gives $\mu^*=44.6\pm 1.8$ Pa, demonstrating the existence of an instability threshold at $\mu^*$ (Fig.\ref{fig : fit grid}-c). This threshold corresponds to a critical dimensionless acceleration $\alpha^*=\frac{\rho g h}{\mu^*}=6.05\pm 0.25$.  
\\

\begin{figure}[!h]
\includegraphics[width=0.5\textwidth]{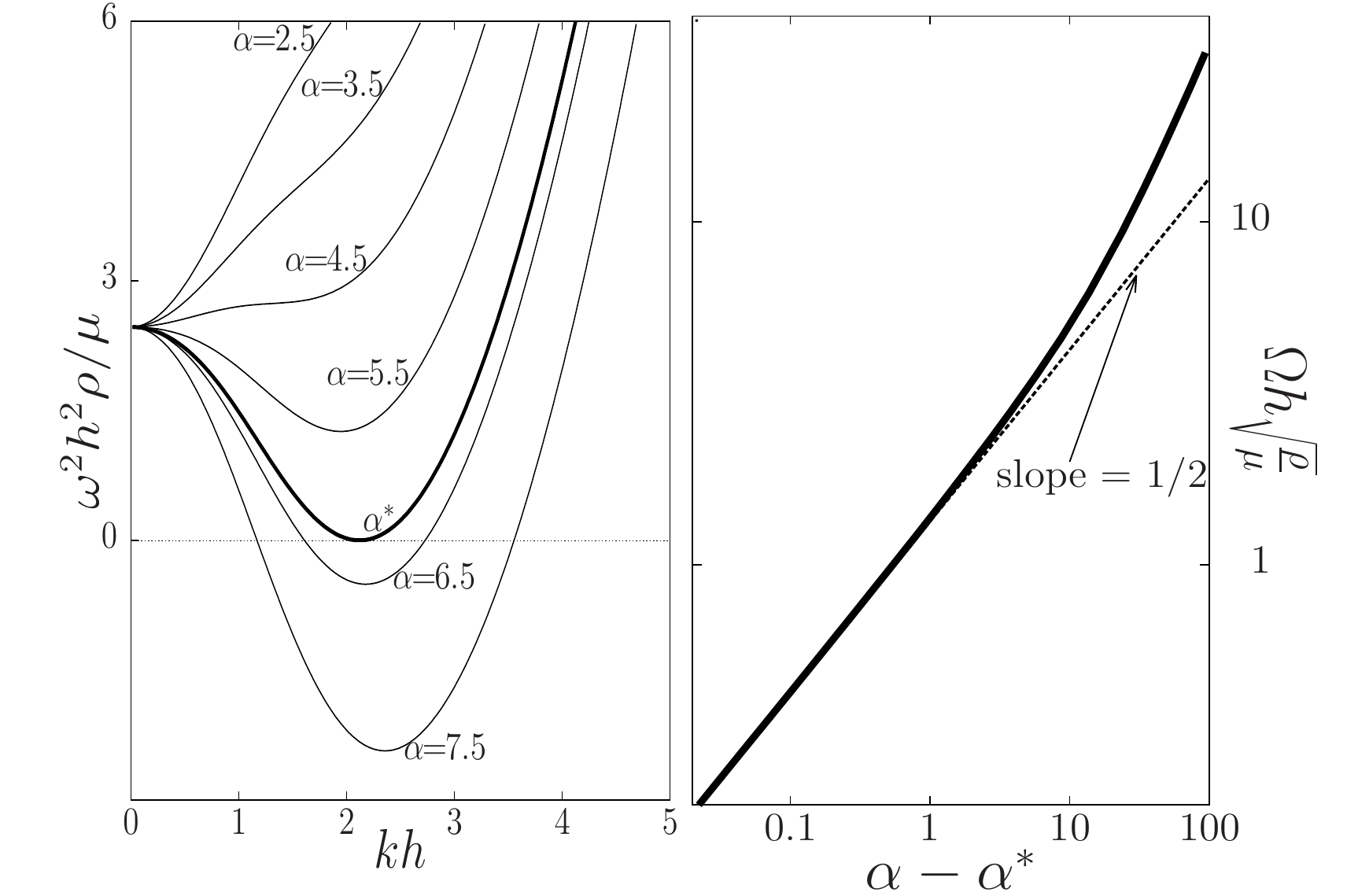}
\caption{Predictions for waves propagating at the surface of a heavy elastic material. {\bf Left:} dimensionless frequency squared $\omega^2 h^2\mu/\rho$ as a function of the dimensionless wave number $kh$ for different values of $\alpha$. For $\alpha<\alpha^*$, the flat surface is stable and $\omega$ is the Rayleigh frequency. For $\alpha>\alpha^*$  the flat surface is unstable and the growth rate of the most unstable mode is $\Omega=\sqrt{-\omega_{min}^2}$, $\omega_{min}$ being the minimum of $\omega$. {\bf Right:} Dimensionless growth rate of the mode $k$ with the maximum growth rate obtained from Eq.\ref{eqn : dispersion} as a function of the gap with the instability threshold (solid line).}
\label{fig : rate}
\end{figure}

 We expect that the onset of instability will show up when the frequency of a mode of propagation of elastic waves \cite{Rayleigh1885,Bromwich1898,Gilbert1967} at finite wavelength becomes zero. We consider an infinite layer of a heavy and incompressible elastic medium of thickness $h$ with a free downwardly facing surface with air, the other surface being fixed on a rigid substrate.  
We also consider a plane wave propagating in the in-plane ${\hat x}$-direction (see Fig.\ref{fig : fit grid})  with the (small) displacement ${\bf u}={\bf u}(y)e^{i\omega t-kx}$ (boldface being for vectors). Putting this displacement field in the equations of motion for an isotropic and incompressible heavy elastic medium  \cite{landau1981,Kuipers1990} with the boundary conditions described just above, we obtain after linearization a condition for the dimensionless frequency $\tilde{\omega}=\omega h\sqrt{\mu/\rho}$ and the dimensionless wave number $\tilde{k}=kh$ (see Supplement-C):
\begin{equation}
\mbox{det}\left( \begin{array}{cccc}
 0&2\tilde{k}^2&0&\tilde{s}^2+\tilde{k}^2\\
-\tilde{\omega}^2+2\tilde{k}^2&-\alpha \tilde{k}&2\tilde{k}\tilde{s}&-\alpha \tilde{k}\\
\tilde{k}\cosh \tilde{k}&-\tilde{k}\sinh \tilde{k}&\tilde{s}\cosh \tilde{s} & -\tilde{s}\sinh \tilde{s}\\
-\tilde{k}\sinh \tilde{k}&\tilde{k}\cosh \tilde{k}&-\tilde{k}\sinh \tilde{s} & \tilde{k}\cosh \tilde{s}
\end{array}\right)=0,
\label{eqn : dispersion}
\end{equation}
for $\tilde{s}^2=\tilde{k}^2-\tilde{\omega}^2>0$. In Fig.\ref{fig : rate}-left, $\tilde{\omega}$ is plotted from Eq.\ref{eqn : dispersion} as a function of $\tilde{k}$ for various values of $\alpha=\frac{\rho g \mu}{h}$. The curves exhibit a local minimum for $\alpha > 4.5$, resulting in two possible wavelengths for one frequency. The  frequency at the minimum becomes zero for $\alpha=\alpha^*=6.223\cdots$:  the propagation speed of the waves is then zero and a sinusoidal perturbation of the surface with the corresponding  wave number is stationary. The surface is then linearly unstable. The theoretical value of $\alpha^*$ is in good agreement with experimental observations (Fig.\ref{fig : fit grid}). The finite size of our samples therefore has no significant effect on the threshold value, nor the gel-air interfacial tension, which is consistent with the calculation of Supplement-B. Furthermore $\tilde{k}=2.12 $ with $h=2.75$ cm corresponds to 8 cm for the wavelength, consistent with snapshots (b) and (c) of Fig.\ref{fig : module}. 
For  $\alpha>\alpha^*$  we find $\omega^2<0$. Writing $\omega=i\Omega$, we obtain the growth rate $\Omega$ of the  instability (Fig.\ref{fig : rate}-right). We find $\Omega\simeq \sqrt{\frac{1.8\mu}{\rho h^2}}\sqrt{\alpha-\alpha^*}$ for $\alpha\leq 1.3\alpha^*$, corresponding to  a characteristic time ($1/\Omega$) ranging from  0.2 (sample (b) of Fig.\ref{fig : module}) to  0.1 s (sample (g)). Unfortunately, such a  characteristic time cannot be experimentally measured since it is shorter than the duration required to place the sample.

The preceding theoretical study applies for infinitesimally small strains, {\em ie} near the threshold. Well beyond it, the final patterns are obtained after a substantially longer duration (a few seconds for patterns of Fig.\ref{fig : module}-e,f,g).
 A surprising and striking non linear feature of the instability is the difference in the observed patterns between snapshots (f), (g) and (h) of Fig.\ref{fig : module}, all the three corresponding to the same shear modulus with the same container size. The first two are directly obtained from a gel of $40 \pm 0.5$ Pa at room temperature. The third one is obtained after cooling sample of Fig.\ref{fig : module}-(f,g) upside down from 50 degrees down to room temperature. The shear modulus correspondingly decreases from $44\pm 0.5$ Pa to $40\pm 0.5$ Pa. The cooling takes place gradually from the boundaries towards the center of the sample, resulting in shear modulus gradients until thermal equilibrium is reached.  The dramatic difference between  the observed patterns highlights the existence of several equilibrium configurations.
This must be related to a complicated energy landscape with several local minima far from the instability threshold, providing a particularly interesting challenge for non linear physics and morphogenesis. Somehow the notion of instability as introducing a kind of free choice in the evolution of a system shows up here. It implies that the ultimate state reached after such an instability depends not only on the growth of the unstable structure itself but also on uncontrollable or at least hard to control small effects, like various inhomogeneities in space and time. This is clearly evidenced in our experiment. \\ 

We have shown that RTI  exists in real elastic solids and that it can be observed in everyday's gravity field in soft hydrogels. The instability threshold depends on the shear modulus, the thickness and the density of the sample. Our experimental set-up with soft elastic gels has enabled a quantitative comparison with a linear theory.  This has allowed us to identify the basic ingredients of this instability and the way it appears.  Measuring the dispersion relation of surface waves can be a mean to detect the proximity of the threshold, and therefore to predict an impending change. 

These results open the way for further fundamental studies, for instance concerning the dynamic formation and the large-scale organization of the patterns, which are both of great importance for non linear physics and morphogenesis.
RTI in solids should  also be found in more complex situations, such as biology, geology and industrial processing, with visco-plastic, visco-elastic or non-isotropic materials. Moreover, the instability is expected to occur in more extreme conditions (high accelerations, strong and non-uniform gravitational fields) where the direct observation is hardly possible. We believe that our work lays foundation to address such more complex cases.\\

The authors are indebted to E. Bouchaud and M. Destrade for interesting discussions. This work has been supported by ANR under Contract no. ANR-2010-BLAN-0402-1 (F2F).\\

\newpage
{\bf Supplementary material}\\
\renewcommand{\thesection}{\Alph{section}-}
\setcounter{section}{0} 
\section{Measurement of the shear modulus of soft polyacrylamide gels.}

\begin{figure}[!h]
\begin{center}
\includegraphics[width=0.6\textwidth]{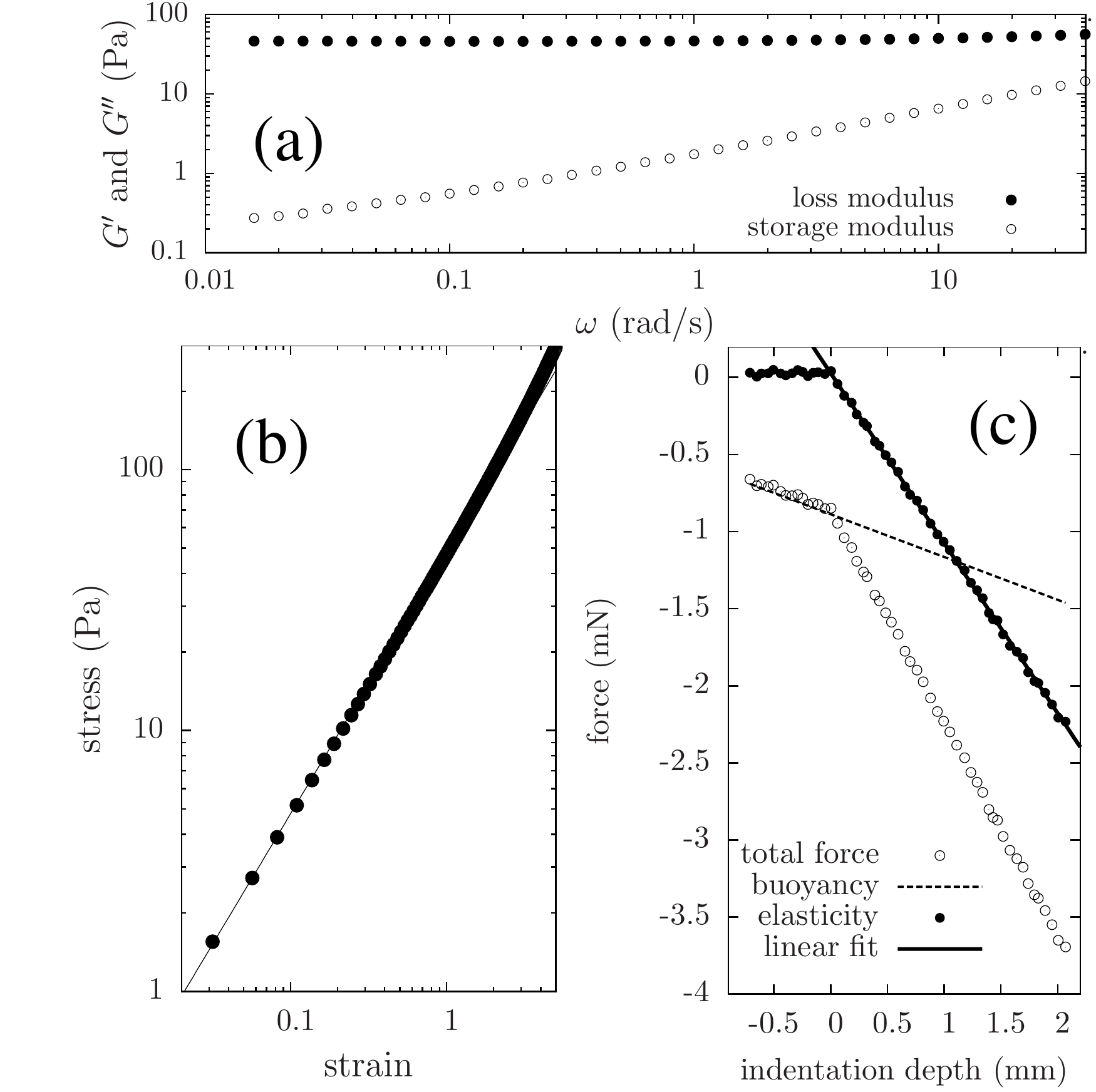}
\end{center}
\caption{ {\bf (a)} Loss and storage moduli. The amplitude of the sinusoidal strain (frequency $2 \pi/\omega$) is 1\%. {\bf (b)} Stress versus shear strain. {\bf (c)} Indentation tests using a flat-ended cylindrical punch, 3 mm diameter. Empty circles : Total normal force applied to a cylindrical indenter as a function of the indentation depth at the surface of a  gel placed in a cylindrical dish facing upwardly and immersed in water. Dashed line : Buoyancy acting on the indenter calculated from the immersed volume. Full circles : Elastic contribution obtained by subtracting buoyancy to the total force. Solid line: linear fit for the elastic contribution leading to a shear modulus of 40 Pa for this gel.}
\label{fig : rheology}
\end{figure}

An accurate measurement of the shear modulus of the materials is crucial for a quantitative analysis of the experimental observations.  
The linear and non linear elastic behaviours of the gels have been checked by classical rheological tests in the cone and plate geometry (using a ARES rheometer, TA Instrument). The storage modulus is much larger than the loss modulus for the relevant timescales of the experiments of the article, showing the elastic linear behaviour of this material (Fig.\ref{fig : rheology}-a). Furthermore, steady stress-sweep tests evidence that the stress is proportional to the strain over a wide range of strains (Fig.\ref{fig : rheology}-b), in agreement with the neo-Hookean model \cite{Ogden1984}.\\
 In situ measurements of the shear moduli are made by indentation tests (Fig.\ref{fig : rheology}-c). In order to remove surface tension contributions \cite{Mora_prl2010,Mora_biot2011}, they are done at the gel-water interface, taking care to proceed fast enough to avoid altering the mechanical properties of the gels due to water diffusion.

\section{Energy minimization including gravity, elasticity and capillarity}

We consider an infinite and heavy elastic layer of height $h$ rigidly fixed under a rigid substrate.  The material is assumed to be homogeneous, incompressible, isotropic, with the density $\rho$ and the shear modulus $\mu$.  It undergoes an outgoing acceleration $g$. The solid-fluid capillary constant is $\gamma$, a physical parameter that is expected to play a role if the order of magnitude of the elasto-capillary length, defined as the ratio of the capillary constant to the shear modulus, is not negligible compared to the other length-scale of the problem, {\em i.e.} the thickness $h$ \cite{Mora_prl2010,Mora_biot2011,Mora2013,Jagota2014}.\\  Here we show that the total energy of this system is stationary against a sinusoidal perturbation of wave number $k$ provided the ratio $\rho g/\mu$ is larger than a given value that depends on $\gamma$ and $k$. We analytically deduce a surface tension dependant threshold for the Elastic Rayleigh Taylor Instability and the corresponding wave number. \\

We limit our exposition to the 2D case. A deformation is characterized by a map from the undisturbed state with coordinates $(x,y)$ to a disturbed state ${\bf{R}}(x,y)= (X(x, y), Y(x, y))$ (boldface being here and thereafter for vectors). In this Lagrangian framework, the actual location in space is parametrized by the coordinates of the preimage in the undisturbed (rest) state.  \\
Incompressibility is imposed by writing that the determinant of the first derivatives of ${\bf{R}}(x,y)$ is equal to one, namely that 
\begin{equation}X_{,x} Y_{,y} - X_{,y} Y_{,x} = 1,\label{eqn : incompressibilite} \end{equation}
where $X_{,x}$ is for $\frac{\partial X}{\partial x}$, etc.\\
\begin{figure}[!h]
\begin{center}
\includegraphics[width=0.45\textwidth]{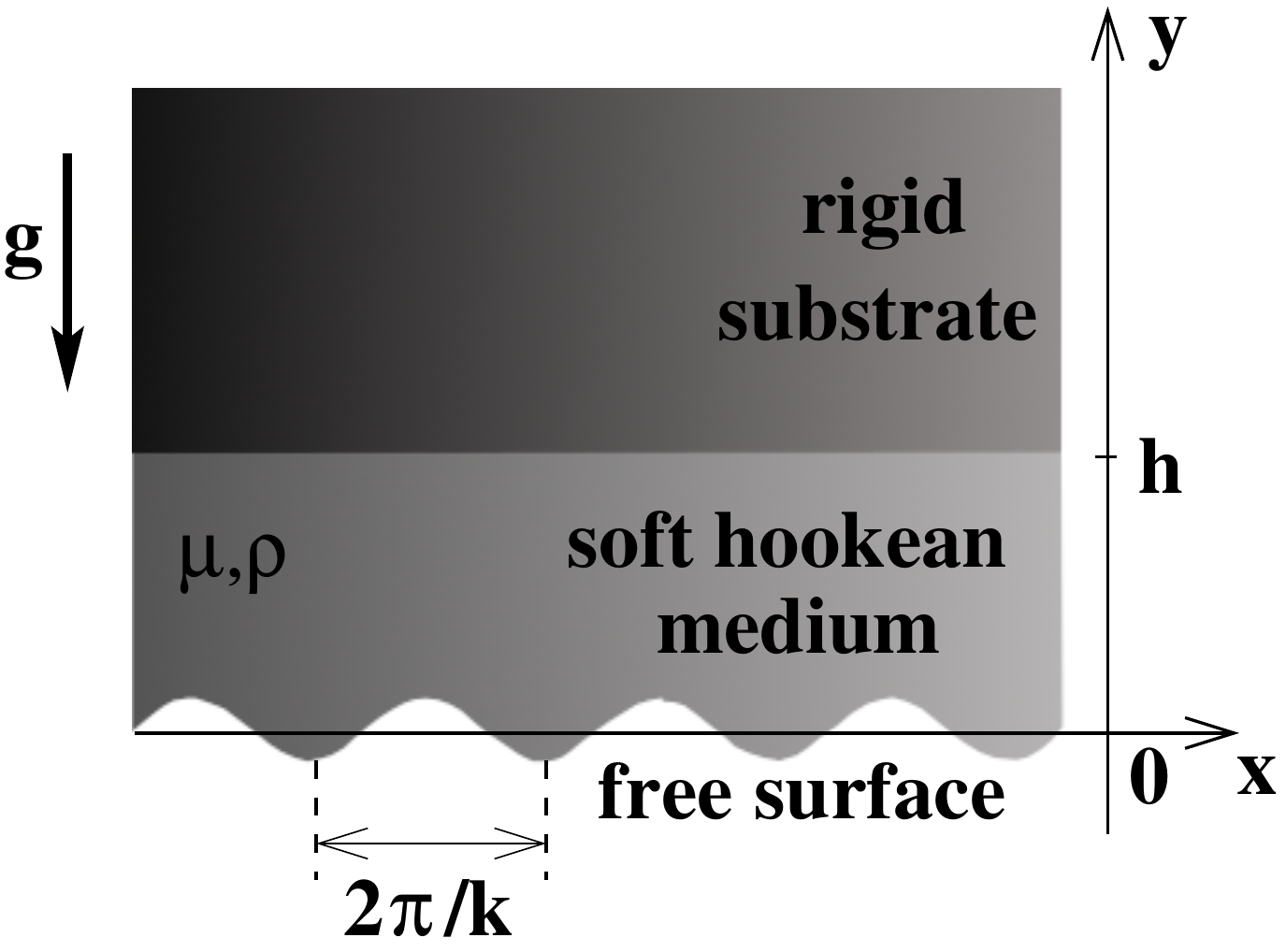}
\end{center}
\caption{Scheme of a sinusoidal perturbation at the free surface of an elastic layer of height $h$ subjected to an acceleration $g$.  The surface at $y=h$ is fixed.} \label{fig : schema onde}
\end{figure} 
Let $Y=\zeta(X)$ be the equation of the surface in the disturbed state. The boundary condition (b.c.) on the surface is the condition that the pair $(X_{\partial} (x) = X(x,0), Y_{\partial} (x) = Y(x,  0))$ is a parametric representation of the curve of equation $ Y_{\partial} = \zeta(X_{\partial})$, the subscript $\partial$ being to indicate that a quantity is evaluated on the free surface ($y=0$, Fig.\ref{fig : schema onde}).\\
Assuming a capillary energy  proportional to  $\gamma$, the energy of the system reads:
\begin{equation}
{\cal E}=\int \mbox{d}x\mbox{d}y \left({\cal W} +\rho g y \right)+\gamma \int \mbox{d}x \sqrt{X^2_{\partial,x}+Y^2_{\partial,x}},
\end{equation}
with ${\cal W}$ the strain energy density function of the elastic material \cite{Ogden1984}. 
Imposing zero variation of $\mathcal{E}$ with respect to $X(.)$ and $Y(.)$ yields the equilibrium equations, both in the bulk (Cauchy-Poisson equations) and at the surface (boundary conditions). The general equations have been previously \cite{wedgeCap} established for the neo-Hookean elasticity, known to describe fairly well the soft solids, or hydrogels, used for experiments. It is the natural extension of Hooke's law and it does not introduce any new physical parameter. From \cite{wedgeCap}, the Cauchy-Poisson equations reads  
\begin{eqnarray}
\frac{\partial}{\partial x}\left( \mu X_{,x}-qY_{,y}\right)+\frac{\partial}{\partial y}\left( \mu X_{,y}+qY_{,x}\right)&=&0, \label{eqn : Cauchy neo 1}\\
\frac{\partial}{\partial x}\left( \mu Y_{,x}+qX_{,y}\right)+\frac{\partial}{\partial y}\left(\mu Y_{,y}-qX_{,x}\right)&=&\rho g, \label{eqn : Cauchy neo 2}
\end{eqnarray}
and the boundary conditions at $y=0$ are:
\begin{eqnarray}
\mu X_{,y}+qY_{,x}
-\gamma \frac{\partial}{\partial x}
\left(\frac{X_{\partial,x}}{\sqrt{X_{\partial,x}^2+Y_{\partial,x}^2}} \right)
&=&0, \label{eqn : condition limites capillaire neo 1}\\
\mu Y_{,y}-qX_{,x}
 +\gamma \frac{\partial}{\partial x}
\left(\frac{Y_{\partial,x}}{\sqrt{X_{\partial,x}^2+Y_{\partial,x}^2}} \right)
&=&0, \label{eqn : condition limites capillaire neo 2}
\end{eqnarray}
where the Lagrange multiplier $q(x,y)$ allows to impose the condition $X_{,x} Y_{,y} - X_{,y} Y_{,x} = 1$ everywhere.\\ 
We consider the infinitesimal perturbation of the base state
$$
\left\{ \begin{array}{l}
\displaystyle X(x,y)=x+\varepsilon {\cal X}(y)e^{ikx}\\
\displaystyle Y(x,y)=y+\varepsilon {\cal Y}(y)e^{ikx}\\
\displaystyle q(x,y)=q^{(0)}+\varepsilon q(y)e^{ikx}
\end{array} \right.
\label{eqn : unstressed perturbation}
$$
with $\varepsilon \ll 1$. From Eq.\ref{eqn : Cauchy neo 1}-\ref{eqn : Cauchy neo 2}, $q^{(0)}=C-\rho g y$ (where $C$ is a constant of integration) and Eq.\ref{eqn : condition limites capillaire neo 2} imposes $C=\mu$.\\

From incompressibility (Eq.\ref{eqn : incompressibilite}):
\begin{equation}
ik{\cal X}+{\cal Y}_{,y}=0.
\label{eqn : incompressibilite sinus}
\end{equation}
The Cauchy-Poisson equations (Eq.\ref{eqn : Cauchy neo 1}-\ref{eqn : Cauchy neo 2}) read:
\begin{equation}
-k^2\mu {\cal X}-ikq+\mu {\cal X}_{,yy}-ik \rho g {\cal Y}=0 \Rightarrow \left(\frac{{\cal Y}_{,yyy}}{k^2}-{\cal Y}_{,y}\right)-\frac{\rho g}{\mu}{\cal Y}=\frac{q}{\mu},
\label{eqn : Cauchy inter} 
\end{equation}
$$
-k^2\mu {\cal Y}+\mu {\cal Y}_{,yy}-ik\rho g {\cal X}-q_{,y}=0 \Rightarrow k^2{\cal Y}-{\cal Y}_{,yy}+\frac{\rho g}{\mu}{\cal Y}_{,y}+\frac{q_{,y}}{\mu}=0.
$$
Substituting the Lagrange multiplier, we obtain:
\begin{equation}
 {\cal Y}_{,yyyy}-2k^2{\cal Y}_{,yy}+k^4{\cal Y}=0.
\label{eqn : unstressed equa diff}
\end{equation}

The boundary conditions $y=0$ (Eqs.\ref{eqn : condition limites capillaire neo 1}-\ref{eqn : condition limites capillaire neo 2}) simplify:
\begin{eqnarray}
{\cal X}_{,y}+{\cal Y}_{,x}&=&0, \nonumber \\
{\cal Y}_{,y}-{\cal X}_{,x}-\frac{q}{\mu} &=&-\frac{\gamma}{\mu}{\cal Y}_{,xx}, \nonumber
\end{eqnarray}
and with Eq.\ref{eqn : incompressibilite sinus} one gets for $y=0$:
\begin{eqnarray}
{\cal Y}_{,yy}+k^2{\cal Y}&=& 0, \label{eqn : condition limites capillaire neo 1bis}\\
2{\cal Y}_{,y}-\frac{q}{\mu} &=&\frac{\gamma}{\mu}k^2{\cal Y}. \label{eqn : condition limites capillaire neo 2bis}
\end{eqnarray}
From  Eqs.\ref{eqn : Cauchy inter}  and \ref{eqn : condition limites capillaire neo 2bis} one obtains
\begin{equation}
3{\cal Y}_{,y}-\frac{1}{k^2}{\cal Y}_{,yyy}=\left(-\frac{\rho g}{\mu}+\frac{\gamma}{\mu}k^2\right){\cal Y}.
 \label{eqn : condition limites capillaire neo 2ter}
\end{equation}
A zero horizontal displacement at $y = h$ is imposed because the elastic medium is assumed to be bonded to the substrate. Hence
the boundary conditions at $y=h$ are ${\cal Y}(h)=0$ and ${\cal X}(h)=0$, or equivalently (from Eq.\ref{eqn : incompressibilite sinus})
 \begin{equation} {\cal Y}(h)=0 \mbox{ and } {\cal Y}_{,y}(h)=0.
\label{eqn : bc h}
\end{equation}

The solutions of Eq.\ref{eqn : unstressed equa diff} are of the form ${\cal Y}=(a+by)e^{-ky}+(c+dy)e^{ky}$. Inserting this expression in Eqs.\ref{eqn : condition limites capillaire neo 1bis}-\ref{eqn : condition limites capillaire neo 2ter} (for $y=0$) and Eqs.\ref{eqn : bc h} (for $y=h$) one obtains: 
$$\left\{
\begin{array}{l}
\displaystyle ka-b+kc+d=0\\
\displaystyle \left(\frac{\rho g}{\mu}-\frac{\gamma}{\mu}k^2-2k\right)a  +\left( \frac{\rho g}{\mu}-\frac{\gamma}{\mu}k^2+2k\right)c=0\\
\displaystyle e^{-kh}a+he^{-kh}b+e^{kh}c+he^{kh}d=0\\
\displaystyle -ke^{-kh}a+(1-kh)e^{-kh}b+ke^{kh}c+(1+kh)e^{kh}d=0.
\end{array} \right.
$$
This system has a nonzero solution $(a,b,c,d)$ for: 
$$
\left| \begin{array}{cccc}
1&-1&1&1\\
\left(\frac{\rho g}{\mu k}-\frac{\gamma k}{\mu}-2\right)&0&\left(\frac{\rho g}{\mu k}-\frac{\gamma k}{\mu}+2\right)&0\\
e^{-kh}&khe^{-kh}&e^{kh}&khe^{kh}\\
-e^{-kh}&(1-kh)e^{-kh}&e^{kh}&(1+kh)e^{kh}
\end{array} \right| = 0,
$$
which simplifies in
\begin{equation}
\left(\frac{\gamma}{\mu h}\right)\left(-e^{2\tilde{k}}+4\tilde{k}+e^{-2\tilde{k}}\right)\tilde{k}+\left(\frac{\rho g h}{\mu}\frac{1}{\tilde{k}}-2\right)e^{2\tilde{k}}-4\left(\frac{\rho g h}{\mu}+1+2\tilde{k}^2\right)-\left(2+\frac{\rho g h}{\mu}\frac{1}{\tilde{k}}\right)e^{-2\tilde{k}}=0,
\end{equation}
or equivalently, in 
\begin{equation}
\alpha=\frac{2\cosh(2\tilde{k})+2(1+2\tilde{k}^2)+\frac{\gamma}{\mu h}\tilde{k}(\sinh(2\tilde{k})-2\tilde{k})}{\frac{\sinh(2\tilde{k})}{\tilde{k}}-2}
\label{eqn : energy complet}
\end{equation}
with $\tilde{k}=kh$ and $\alpha=\frac{\rho g h}{\mu}$. The flat interface is neutrally stable against an infinitesimal perturbation of dimensionless wave number $\tilde{k}$ for a dimensionless acceleration $\alpha=\frac{\rho g h}{\mu}$ given by the condition of Eq.\ref{eqn : energy complet}. For $\alpha$ larger than the value calculated from Eq.\ref{eqn : energy complet} the flat interface is unstable against an infinitesimal perturbation of dimensionless wave number $\tilde{k}$. 

\begin{figure}[!h]
\begin{center}
\includegraphics[width=0.6\textwidth]{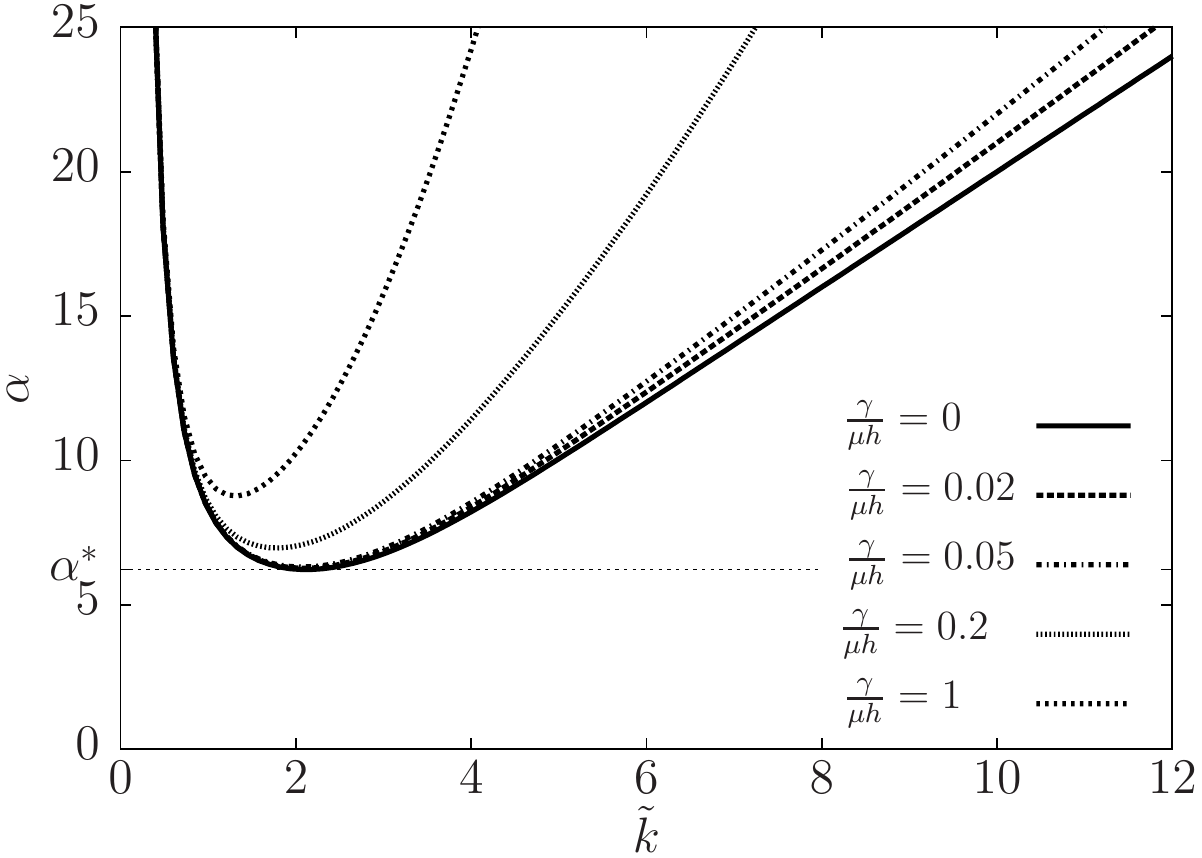}
\end{center}
\caption{Bold line: Dimensionless acceleration ($\alpha=\frac{\rho g h}{\mu}$) so that the flat interface is neutrally stable against an infinitesimal sinusoidal perturbation of dimensionless wave number $\tilde{k}$,  according to Eq.\ref{eqn : energy complet}. Its minimum value $\alpha^*\simeq 6.223$ corresponds to the instability threshold.} \label{fig : energy complet}
\end{figure}

Let us first suppose that $\gamma/\mu\ll h$. In this case
\begin{equation}
\alpha=\frac{2\tilde{k}\left(2\tilde{k}^2+\cosh(2\tilde{k})+1\right)}{\sinh(2\tilde{k})-2\tilde{k}}.
\label{eqn : energy}
\end{equation}
Plotted as a function of $\tilde{k}$, $\alpha$ reaches an absolute minimum for $\tilde{k}=\tilde{k}^* \equiv 2.12\cdots$. It is then equal to $\alpha=\alpha^*\equiv 6.223\cdots$ (see the continuous bold line of Fig.\ref{fig : energy complet} which corresponds to $\gamma=0$).  The flat surface is unstable against an infinitesimal perturbation of any wavelength for any dimensionless acceleration larger than $\alpha^*$. Otherwise the flat interface is stable (with respect to infinitesimal perturbations). $\alpha^*$ defines the instability threshold. The instability first appears (at the threshold) for a wavelength equal to $2h\pi/\tilde{k}^*\simeq 2.96h$. \\   

The minimum of $\alpha$ depends on the surface tension: its value increases as $\frac{\gamma}{\mu h}$ is larger, and the corresponding value of $\tilde{k}$ is smaller (Fig.\ref{fig : energy complet}). As expected, the surface tension stabilizes the interface, an effect which is more pronounced as the wavelengths are shorter since the most unstable wave number is shifted to the smaller values when the surface tension is increased.\\

The polyacrylamide gel-air interfacial tension is of the order of magnitude of the water-air surface tension as the gel mainly consists of water. Therefore, $\gamma\sim 70~mN/m$ in our experiments. The sample thickness being $h\sim 3~cm$ and the measured modulus at the threshold $\mu \sim 45~Pa$, one obtains $\frac{\gamma}{\mu h}\sim 0.05$. From Eq.\ref{eqn : energy complet}, the shift of $\alpha$ at the threshold is not expected to be significant (see Fig.\ref{fig : energy complet}). Therefore surface tension effects are not expected to play a role in our experiments at the instability threshold.\\

 Note that surface tension effects should be observable with thinner samples. For instance, consider a sample of thickness $h\sim 0.65~cm$ (made of the same gel with $\rho=10^3~kg/m^3$, $g=9.8~m.s^{-2}$, $\gamma\sim 70~mN/m$). Taking $\rho g h/\mu=\alpha^*$ yields $\mu \sim 10$ and $\frac{\gamma}{\mu h}\sim 1$. Fig.\ref{fig : energy complet} indicates that the threshold is shifted by the surface tension. This effect should also be observed with harder samples subjected to larger accelerations or with larger density contrasts.

\section{Dispersion relation}

We calculate the dispersion relation of Rayleigh waves propagating at the surface of an infinite and heavy elastic layer of height $h$, and rigidly fixed under a rigid substrate. The material is assumed to be homogeneous, incompressible, isotropic, with the density $\rho$ and the shear modulus $\mu$.  It is subjected to an outgoing acceleration $g$. The surface tension is here neglected. We consider here plane waves with infinitely small amplitudes and the model of Hookean elasticity is used.\\

The propagation of elastic surface waves in heavy elastic materials has been addressed by several authors \cite{Bromwich1898,Gilbert1967,Kuipers1990,Vinh2012}. 
These authors had likely in mind the issue of waves propagating on the surface of the Earth. Hence, they were exclusively interested in accelerations directed towards the elastic medium. The effects of gravity on the wave propagation are very small in this case.\\

Here, the calculation  of \cite{Kuipers1990} is modified in accordance with our experimental setup  (outward gravity and no-slip condition, Fig.\ref{fig : schema onde}). The changes in the calculation are highlighted below.\\

The deformation of the elastic medium is characterized by a map from the undisturbed state (flat surface) with coordinates (x,y,z) to a disturbed state ${\bf R}(x,y,z)=(X(x,y,z), Y(x,y,z),Z(x,y,z))$. In the following, we consider a  two-dimensional problem, which amounts to imposing $Z = z$.
The displacement ${\bf u}={\bf R}-{\bf r}$ is assumed to be small, and its Cartesian components are noted $u_x$ (in plane) and $u_y$ (out of plane). Applying the decomposition theorem of Helmholtz we write :
\begin{eqnarray}
u_x&=&\frac{\partial \phi}{\partial x}+\frac{\partial \psi}{\partial y}\\
u_y&=&\frac{\partial \phi}{\partial y}-\frac{\partial \psi}{\partial x}
\end{eqnarray}
where $\phi$ and $\psi$ are functions of $x$, $y$ and $t$. $\phi$ and $\psi$ are  solutions of equations (see Eq.3.3 and Eq.3.5.1 of \cite{Kuipers1990}): 
\begin{equation}
\Delta \phi=0 \mbox{ and } \Delta \psi=\frac{1}{c_2^2}\frac{\partial^2 \psi}{\partial t^2}
\label{eqn : bulk}
\end{equation}
with $c_2=\sqrt{\mu/\rho}$.
Considering a sinusoidal wave of frequency $\omega/2\pi$ and wavelength $k/2\pi$ propagating in the $x$-direction, $\phi$ and $\psi$ can be chosen as (see Eq.3.9 in \cite{Kuipers1990}): 
\begin{eqnarray}
\phi&=&\left(A_1\cosh ky+A_2 \sinh ky\right)e^{i(\omega t-kx)} \label{eqn : phi}\\
\psi&=&\left(B_1\sinh sy +B_2 \cosh sy \right)e^{i(\omega t-kx)}  \label{eqn : psi}
\end{eqnarray}
The boundary conditions at the free surface $y=0$ are (Eqs.3.7 in \cite{Kuipers1990}):
\begin{equation}
\left\{ \begin{array}{l}
2\frac{\partial^2\phi}{\partial x \partial y}+\frac{\partial^2\psi}{\partial y^2}-\frac{\partial^2\psi}{\partial x^2}=0\\
\rho g \left(\frac{\partial \phi}{\partial y}-\frac{\partial \psi}{\partial x}\right)+\rho \frac{\partial^2\phi}{\partial t^2}+2\mu \left(\frac{\partial^2\phi}{\partial y^2}-\frac{\partial^2\psi}{\partial x \partial y}\right)=0
\end{array} \right.
\label{eqn : bc wave 0}
\end{equation}
A zero vertical displacement at $y = h$: $u_y(h)=0$ is imposed. But contrary to \cite{Kuipers1990} where the elastic material could freely slide parallel to the substrate, here a zero horizontal displacement at $y = h$ is also imposed because the elastic medium is assumed to be bonded to the substrate. Hence $u_x(h)=0$ and Eqs.3.8 are replaced by:
\begin{equation}
\left\{\begin{array}{l}
\left.\frac{\partial \phi}{\partial x}\right|_{y=h}+\left.\frac{\partial \psi}{\partial y}\right|_{y=h}=0\\
\left. \frac{\partial \phi}{\partial y}\right|_{y=h}-\left.\frac{\partial \psi}{\partial x}\right|_{y=h}=0
\end{array}\right.
\label{eqn : bc wave h}
\end{equation}

Substituting the expressions for $\phi$ (Eq.\ref{eqn : phi}) and $\psi$ (Eq.\ref{eqn : psi}) in the boundaries conditions (Eqs.\ref{eqn : bc wave 0}-\ref{eqn : bc wave h}) yields:
\begin{equation}
\left\{ \begin{array}{l}
2ik^2A_2+(s^2+k^2)B_2=0\\
\rho g\left(-kA_2+ikB_2\right)-\rho \omega^2 A_1+2\mu\left(k^2A_1-iksB_1\right)=0\\
-ik\left(A_1\cosh kh+A_2 \sinh kh\right)-s\left(B_1\cosh sh +B_2 \sinh sh \right)=0\\
-k\left(A_1\sinh kh+A_2 \cosh kh\right)+ik\left(B_1\sinh sh +B_2 \cosh sh \right)=0
\end{array} \right.
\label{eqn : wave system}
\end{equation}
Following \cite{Kuipers1990} we introduce the dimensionless parameters:
$$
\tilde{k}=kh,~\tilde{\omega}=\frac{h}{c_2}\omega,~\alpha=\frac{\rho g h}{\mu},~\tilde{s}=hs
$$
The condition for the linear system (Eq.\ref{eqn : wave system}) has nonzero solutions, {\em ie} for a wave with a wave number $k$ and a frequency $\omega/2\pi$ can propagate, is:
$$
\left| \begin{array}{cccc}
0&-2i\tilde{k}&0&\left(\tilde{s}^2+\tilde{k}^2\right)\\
\left(2\tilde{k}^2-\tilde{\omega}^2\right) & -\alpha \tilde{k} & 2i\tilde{k}\tilde{s} & -i\alpha \tilde{k}\\
-i\tilde{k}\cosh \tilde{k}&i\tilde{k}\sinh \tilde{k}& \tilde{s}\cosh \tilde{s}&-\tilde{s}\sinh \tilde{s}\\
-\tilde{k}\sinh \tilde{k} & \tilde{k} \cosh \tilde{k} & -i\tilde{k}\sinh \tilde{s} & i\tilde{k}\cosh \tilde{s}
\end{array}\right|=0
$$

From Eq.\ref{eqn : bulk}, $s^2=k^2-(\omega/c_2)^2$. Hence:
$$
\left| \begin{array}{l}
\tilde{s}=\sqrt{\tilde{k}^2-\tilde{\omega}^2} \mbox{ if } \tilde{k}>\tilde{\omega}\\
\tilde{s}=i\sqrt{\tilde{\omega}^2-\tilde{k}^2}=i\tilde{s}' \mbox{ if } \tilde{k}<\tilde{\omega}
\end{array} \right.
$$

If $\tilde{k}>\tilde{\omega}$ the propagation condition writes:
\begin{equation}
\left| \begin{array}{cccc}
  0&2\tilde{k}^2&0&\tilde{s}^2+\tilde{k}^2\\
-\tilde{\omega}^2+2\tilde{k}^2&-\alpha \tilde{k}&2\tilde{k}\tilde{s}&-\alpha \tilde{k}\\
\tilde{k}\cosh \tilde{k}&-\tilde{k}\sinh \tilde{k}&\tilde{s}\cosh \tilde{s} & -\tilde{s}\sinh \tilde{s}\\
-\tilde{k}\sinh \tilde{k}&\tilde{k}\cosh \tilde{k}&-\tilde{k}\sinh \tilde{s} & \tilde{k}\cosh \tilde{s}
\end{array} \right|=0
\label{eqn : dispersion 1}
\end{equation}

If $\tilde{k}<\tilde{\omega}$:
\begin{equation}
\left| \begin{array}{cccc}
0&2\tilde{k}^2&0&\tilde{k}^2-\tilde{s}'^2\\
-\tilde{\omega}^2+2\tilde{k}^2&-\alpha \tilde{k}&2\tilde{k}\tilde{s}'&-\alpha \tilde{k}\\
\tilde{k}\cosh \tilde{k}&-\tilde{k}\sinh \tilde{k}&\tilde{s}'\cos \tilde{s}'&\tilde{s}'\sin \tilde{s}'\\
-\tilde{k}\sinh \tilde{k} & \tilde{k}\cosh \tilde{k}&-\tilde{k}\sin \tilde{s}' & \tilde{k} \cos \tilde{s}'
\end{array} \right|=0
\label{eqn : dispersion 2}
\end{equation}
Eqs.\ref{eqn : dispersion 1}  and \ref{eqn : dispersion 2} together define the dispersion relation of elastic surface waves propagating in a heavy material of height $h$. Plotting $\omega^2$ as a function of $k$ evidences that
\begin{itemize}
\item{(i) for $-\infty<\alpha <4.5$, $\omega^2(k)$ is an increasing function,}
\item{(ii) for $4.5\cdots <\alpha <6.223\cdots$, $\omega^2(k)$ has a local minimum that is positive,}
\item{(iii) for $\alpha= 6.223\cdots$, $\omega^2(k)$ vanishes for a given value of k. This corresponds to the instability threshold,}
\item{(iv) for $\alpha >6.223\cdots$, $\omega^2(k)$ is negative in a range of $k$. The related modes are unstable and $\sqrt{-\omega^2(h)}$ is their growth rate.}

Note that taking $\tilde{\omega}=0$ in Eq.\ref{eqn : dispersion 1} (and hence $\tilde{s}=\tilde{k}$) yields Eq.\ref{eqn : energy}.

\end{itemize}

\end{document}